\documentclass{aa}  

\usepackage{graphicx}
\usepackage{txfonts}
\begin{document} 

   \title{Can supernova shells feed supermassive black holes in galactic nuclei?}

   \author{J. Palou\v s
          \inst{1}
          \and
          S. Ehlerov\' a
          \inst{1}
          \and
          R. W\" unsch
          \inst{1}
          \and
          M. R. Morris
          \inst{2}
          }

   \institute{Astronomical Institute, Academy of Sciences,  
              Bo\v{c}n\'{\i} II 1401, Prague, Czech Republic\\  
              \email{palous@ig.cas.cz}    
         \and
             Department of Physics and Astronomy, University of California
             Los Angeles, CA 90095-1547, USA\\
             \email{morris@astro.ucla.edu}
             }

   \date{Received June 26, 2020; accepted October 12, 2020}

 
  \abstract
  {}
  {We simulate shells created by supernovae expanding into the interstellar medium (ISM) of the nuclear region of a galaxy, and analyze how the shell evolution is influenced by the supernova (SN) position relative to  the galactic center, by the interstellar matter (ISM
    ) density, and by the combined gravitational pull of the nuclear star cluster (NSC) and supermassive black hole (SMBH).}
   {We adopted simplified hydrodynamical simulations using the infinitesimally thin layer approximation in 3D (code RING) and determined whether and where the shell expansion may bring new gas into the inner parsec around the SMBH.}
   {The simulations show that supernovae occurring within a conical region around
the rotational axis of the galaxy can feed the central accretion disk
surrounding the SMBH.  For ambient densities between 10$^3$ and 10$^5$
cm$^{-3}$, the average mass deposited into the central parsec by individual
supernovae varies between 10 to 1000 solar masses depending on the ambient
density and the spatial distribution of supernova events.   Supernovae
occurring in the aftermath of a starburst event near a galactic center can
supply two to three orders of magnitude more mass into the central parsec,
depending on the magnitude of the starburst.  The deposited mass typically
encounters and joins an accretion disk.  The fate of that mass is then
divided between the growth of the SMBH and an energetically driven outflow
from the disk.} 
   {}
    \maketitle
%

\section{Introduction}

The mechanisms leading to the formation and growth of SMBHs
in the centers of their host galaxies is an open unsolved problem in astrophysics. The mass of the SMBH is correlated with the mass of the bulge \citep{2013ARA&A..51..511K} and some extraordinary starburst activities of active galactic nuclei (AGN) may be connected to accreting SMBHs \citep{2020arXiv200712026Y}. However, this starburst activity usually occurs in regions enshrouded by dust, hiding completely their emission in optical wave bands. The energy is radiated away usually at infrared wavelengths and galaxies with abundant accretion onto a central SMBH appear as ultra luminous infrared galaxies (ULIRG) \citep{1996ARA&A..34..749S}. Interstellar matter can be funneled into central regions of galaxies by mergers and interactions of gas-rich galaxies that trigger the starburst activity. The feedback driven by young and massive stars  leads to outflows \citep{2014ApJ...792..101D}, thereby reducing the interstellar matter (ISM)
supply to the central SMBH, limited also by the conservation of angular momentum.

Nuclear star clusters (NSCs)
are found nestled in the cores of galaxies with SMBHs at the bottom of the galaxy's potential wells. Their typical half-mass radii are (2 - 5) pc, with the full extent going out to (5 - 50) pc and estimated masses are in the range (10$^5$  - 10$^9$) M$_\odot$ \citep{2017ApJ...849...55S}. NSC masses appear to be related to masses of their host galaxies sharing a relationship similar to that between  SMBH masses and galactic bulge masses \citep{2013ARA&A..51..511K}. The formation of a NSC follows two possible scenarios. The first assumes that globular clusters formed in the galaxy outside of the galactic center and then spiralled down into the galaxy's core because of dynamical friction before merging there to form the NSC \citep{1975ApJ...196..407T}.
However, it appears that there are not enough RR Lyrae stars in the NSC of our Galaxy to be consistent with this first scenario: if the NSC were formed from globular clusters, the expected number of  RR Lyrae stars would be almost an order of magnitude higher compared to numbers identified by the Hubble space telescope 
observations of the center of the Milky Way \citep{2017MNRAS.471.3617D}.
The alternative scenario proposes in situ formation of the NSC developing from gas migrating into the galactic center due to internal processes in the galactic disc or as a result of interactions with other galaxies \citep{2008JPhCS.131a2044S}.

We investigate here the evolution of shells expanding into the interstellar medium surrounding the SMBH in the centers of galaxies inside of NSCs.   The expanding shells are the result of energy and momentum inserted by young and   
massive stars of the NSC in the form of winds, radiation, and supernova explosions. The evolution of shells expanding into the interstellar medium resulting from either supernovae or stellar winds 
has been described in many papers and books, for example, by \citet{ostriker1988} and \citet{bisnovatyj-kogan1995}. 
Here we summarize very briefly how the blast wave resulting from a supernova 
explosion evolves. 
The first stage is the phase of free expansion 
followed by the so-called Sedov-Taylor phase characterized by  constant 
thermal energy inside the blast wave driving the expansion. 
When the radiative losses in the shell of the swept-up gas become important 
and the heat energy created by the compression of interstellar gas is
radiated away, the dense wall of the 
shell behind the shock shrinks to a thin layer and the structure 
enters the pressure-driven, thin-shell phase. Later, when the pressure inside the shell drops, 
the wave is no longer driven by the interior pressure and it only keeps its 
momentum. This is the so-called snowplow phase. Even later, when the expansion velocity decreases below the local sound speed, the mass accumulation ceases and the thin shell dissolves.  

The main intention of this paper is to investigate whether supernovae (SNe)
occurring in NSCs can deliver mass into the central region inside of the circumnuclear disk surrounding the SMBHs in the centers of galaxies. We explore the importance of the SN position relative to the SMBH and we analyze how the inserted energy, mass, and the density of the interstellar medium influence the mass delivery to the SMBH. Here we assume that the gas inside the expanding SN remnant is non-relativistic (gas velocity is below a few thousand km s$^{-1}$), and that shells are able to survive encounters with random density fluctuations and stay coherent when they are deformed by the tidal forces of the SMBH and of the NSC. To find suitable energies and positions of SNe for mass delivery near the central SMBH, in this paper we perform  simulations with the simplified hydrodynamical code RING.

In the 3D code RING we assume that the wall of the shell is infinitesimally thin, meaning that its thickness is much smaller than its diameter. In this and future papers, simulations with the fast code RING will map the expanding shell properties for different SNe energies, positions relative to the SMBH and NSC, and densities of the ISM. 
This approach was developed by \citet{kompaneets1960} 
and \citet{bisnovatyj-kogan1982}, and has been used by \citet{tenorio-tagle1987},  
\citet{ehlerova1996}, \citet{silich1996}, \citet{1997A&A...328..121E},
\citet{1999A&A...350..457E}, \citet{2002MNRAS.334..693E}, 
and others.

The paper is organized as follows: in Sects. 2 and 3 we describe the gravitational potential of the SMBH and of the NSC, the density and motion of the ISM, and the position of the supernova relative to the SMBH. The formation of the thin shell, as simulated in 1D with the hydrodynamical code FLASH, is presented in Sect. 4. In Sect. 5 we introduce the 3D code RING, which is used in Sect. 6, where we show the time evolution of the thin shell. Results are given in Sect. 7, and discussed in Sect. 8. Conclusions are shown in Sect. 9.

\section{Gravitational potential of the SMBH and NSC}

   \begin{figure*}
     \centering
     \includegraphics[angle=0,width=0.9\linewidth]{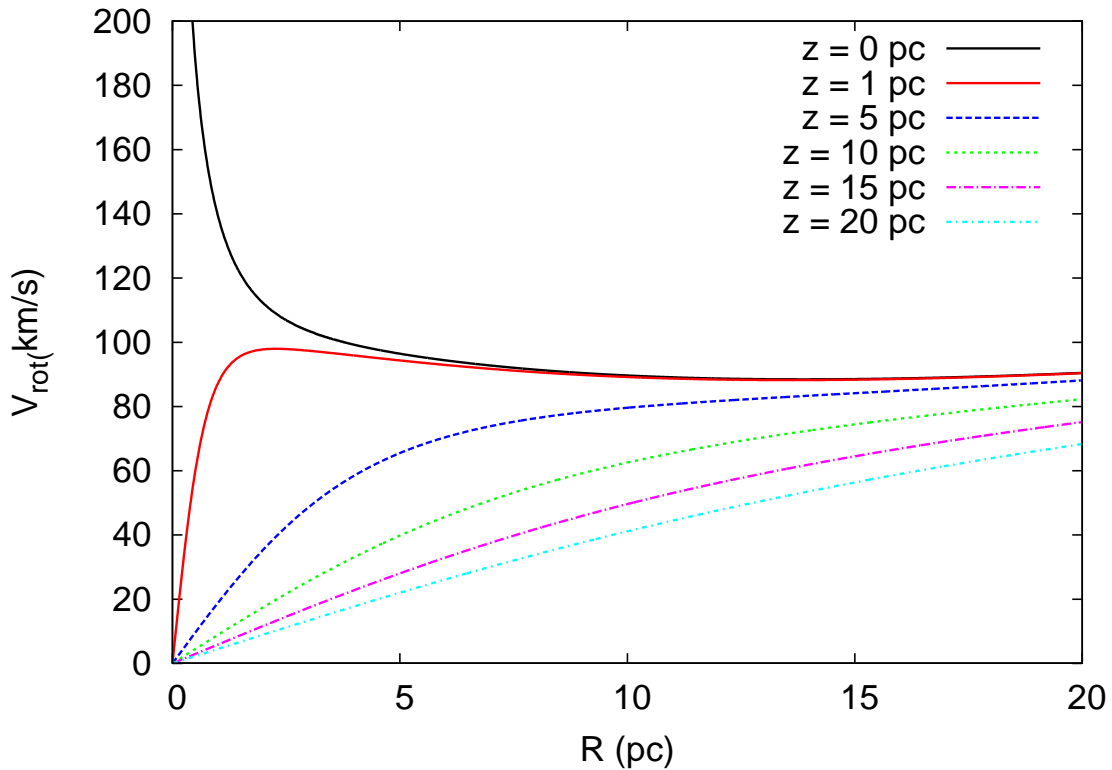} 
     \caption{Interstellar density (ISM)
       rotation velocity, $V_{\rm rot}$, is shown as a function of $R$ for different distances $z$ from the symmetry plane.} 
   
              \label{rotation}
    \end{figure*}
The gravitational potential of the SMBH, $\Psi_{\rm SMBH}$, is given as
\begin{equation}  
\Psi_{\rm SMBH} = -{G M_{\rm SMBH} \over (r_{GC} + \epsilon)},   
\end{equation}
where $G $ is the gravitational constant, $M_{\rm SMBH}$ is the SMBH mass,
and $r_{GC}$ is the distance from the galactic center. Here we assume that the SMBH is located exactly at the Galactic center and that the NSC is similarly centered there. A small constant $\epsilon$ has
a value of $10^{-6}$ pc preventing infinite values of the
potential and of its derivatives. The SMBH mass, $M_{\rm SMBH}$,
is a
free parameter with a value in the interval (10$^6$ - 10$^{9}$) M$_\odot$.
In this paper, we keep $M_{\rm SMBH}$ constant,
since we assume the mass delivered to it during the interaction is much less than the mass $M_{\rm SMBH}$ itself.
In the case
of the SMBH in the center of the Milky Way, we take
$M_{\rm SMBH} = 4 \times 10^6$ M$_\odot$.

The stellar density distribution in the NSC, $\rho_{\rm NSC}$, is given as
a combination of two $\gamma$-models proposed by \citet{1993MNRAS.265..250D}
in the form suggested by \citet{2015MNRAS.447..948C}:
\begin{equation}
\rho_{\rm NSC} = \rho_{\rm NSC_1} + \rho_{\rm NSC_2},
\end{equation}  

\begin{equation}
  \rho_{\rm NSC_i}(r_{GC}) = {{3 - \gamma_i} \over {4 \pi}} {{M_{\rm NSC_i} a_i} \over {r_{GC}^{\gamma_i} (r_{GC} + a_i)^{4 - \gamma_i}}},
  \label{NSC-density}
\end{equation}
where $i = 1,2$ and $M_{\rm NSC_i}$ is the mass of the two individual nuclear
stellar subclusters.

For simulations of the shell expanding in the NSC of the Milky Way, following \citet{2015MNRAS.447..948C} we adopt 
\begin{eqnarray}
M_{\rm NSC_1} = 2.7 \ 10^7 M_\odot, \gamma_1 = 0.51, a_1 = 3.9 \ {\rm pc};\\
M_{\rm NSC_2} = 2.8 \ 10^9 M_\odot, \gamma_2 = 0.07, a_2 = 94.4 \ {\rm pc}.
\end{eqnarray}
The total NSC mass is $M_{\rm NSC} = M_{\rm NSC_1} + M_{\rm NSC_2}$. The mass
$\rm NSC_2$
appears to coincide with the one described by \citet{1996Natur.382..602S}.
In future simulations,  the division between the two stellar subclusters will be varied 
together with the total value of $M_{NSC}$: 
(10$^5 < M_{\rm NSC} <  10^9$) M$_\odot$.

The corresponding gravitational potential of the NSC is
\begin{equation}
\Psi_{\rm NSC} (r_{GC}) = -\sum_{i=1}^2 {{G M_{\rm NSC_i}} \over a_i} {1 \over {(2 - \gamma_i)}} \left( 1 - ({r_{GC} \over {r_{GC} + a_i}})^{2 - \gamma_i}\right).
\end{equation}

In this paper we neglect the contribution of the ISM to the overall
gravitational field, and the total gravitational potential, $\Psi (r_{GC})$, is given as the combination of the SMBH and NSC only:
\begin{equation}
\Psi (r_{GC}) = \Psi_{\rm SMBH} + \Psi_{\rm NSC}. 
\end{equation}

\section{Density and motion of the ISM and positions of the supernova}

In this initial exploration of how a supernova will affect gas dynamics near a
supermassive black hole, we simplify the physical conditions by assuming that
the ambient ISM is distributed in a homogeneous way with a constant density, $n_{out}$, which is a free parameter. We explore three values: $n_{out} = 10^3, 10^4$, and 10$^5$ cm$^{-3}$ \citep{2015ApJ...812...72A,2020A&ARv..28....4N}. In subsequent communications we shall also analyze shells expanding in lower or higher ISM densities.

The ambient ISM rotates in the gravitational field of the
SMBH and NSC with the plane of rotation parallel to the symmetry plane of the hosting galaxy.
The circular rotation velocity, $V_{\rm rot}(R, z)$, is given as
\begin{equation}
  V_{\rm rot} (R, z) = \left({{d \Psi(r_{GC})} \over {dr_{GC}}} r_{GC} \right)^{1/2} {{R} \over {r_{GC}}},
  \label{eq-rotation}
\end{equation}
where $r_{GC} = \sqrt{R^2 + z^2}$ is the galactocentric distance and $(R, z)$
are the galactocentric cylindrical
coordinates with the rotational axis perpendicular to the $z = 0$ plane.
We assume that in the $z = 0$ plane the rotation of the unperturbed gas completely balances the
gravitational attraction. At the distance $z$ from the plane, rotation
slows down since it only balances the component of gravitational force parallel to the plane of rotation; however, the ISM density is kept homogeneous and independent of the distance $z$.
The rotational velocity as a function
of cylindrical coordinates $(R, z)$ is given in Fig. \ref{rotation}.

We analyze the expansion of SNe shells from different initial positions within the NSC
close to the SMBH. The initial values $(R_0, z_0)$  are varied inside the volume
\begin{equation}
 1 < r_{GC} < 150 pc.
\end{equation}
In subsequent papers we shall also consider the SNe motion relative to the ISM at the instant of the SNe explosion.

\section{Formation of the thin shell} 

Over a few second, a supernova injects energy $E_{tot}$ into the
surrounding ISM creating an expanding shock wave. Here we adopt the
canonical value $E_{tot} = 10^{51}$ erg; other values of $E_{tot}$ will be explored in the future.
To set the initial conditions for the three-dimensional code RING
we first  performed
one-dimensional simulations with the grid-based hydrodynamical code FLASH v4.3  \citep{2000ApJS..131..273F} using the Piecewise Parabolic Method \citep{1984JCoPh..54..174C} with the time-step controlled by the Courant-Friedrichs-Lewy criterion and  with equilibrium radiative
cooling using the prescription of \citet{2009A&A...508..751S}.

The simulation starts from a small spherical volume
into which we inserted SNe ejecta of mass $M_{ej}$ with total energy $E_{tot}$. The initial conditions of the free-expansion phase are given by  \citet{2017MNRAS.465.3793T} with the density $\rho_{ej}$ and velocity $v_{ej}$ profiles specified as
\begin{equation}  
\rho_{\rm ej} = {1 \over 4 \pi} {M_{ej} \over r_{ej}^3} \left( {r_{ej} \over r}\right)^2   
\end{equation}
\begin{equation}  
v_{ej} = \left(6{E_{SN} \over M_{ej}}\right)^{1/2} \left({r \over r_{ej}}\right),
\end{equation}
 where $r$ is the distance from the SN explosion, and $r_{ej}$ is the radius of the spherical volume where the ejecta are distributed. Here
 we take $M_{\rm ej} = 10 $ M$_\odot$ and $r_{ej} = 0.24$ pc.
 The shock wave expands into a warm ($T = 10^4$ K) ambient  medium of density  $\rho_{out} = \mu \  n_{out}$ with $\mu = 0.609 \ m_H$, where $m_H$ is the atomic mass unit, and $n_{out} = 10^3, 10^4,$ or $10^5$ cm$^{-3}$.

\begin{figure*}
\centering 
\includegraphics[angle=0,width=0.49\linewidth]{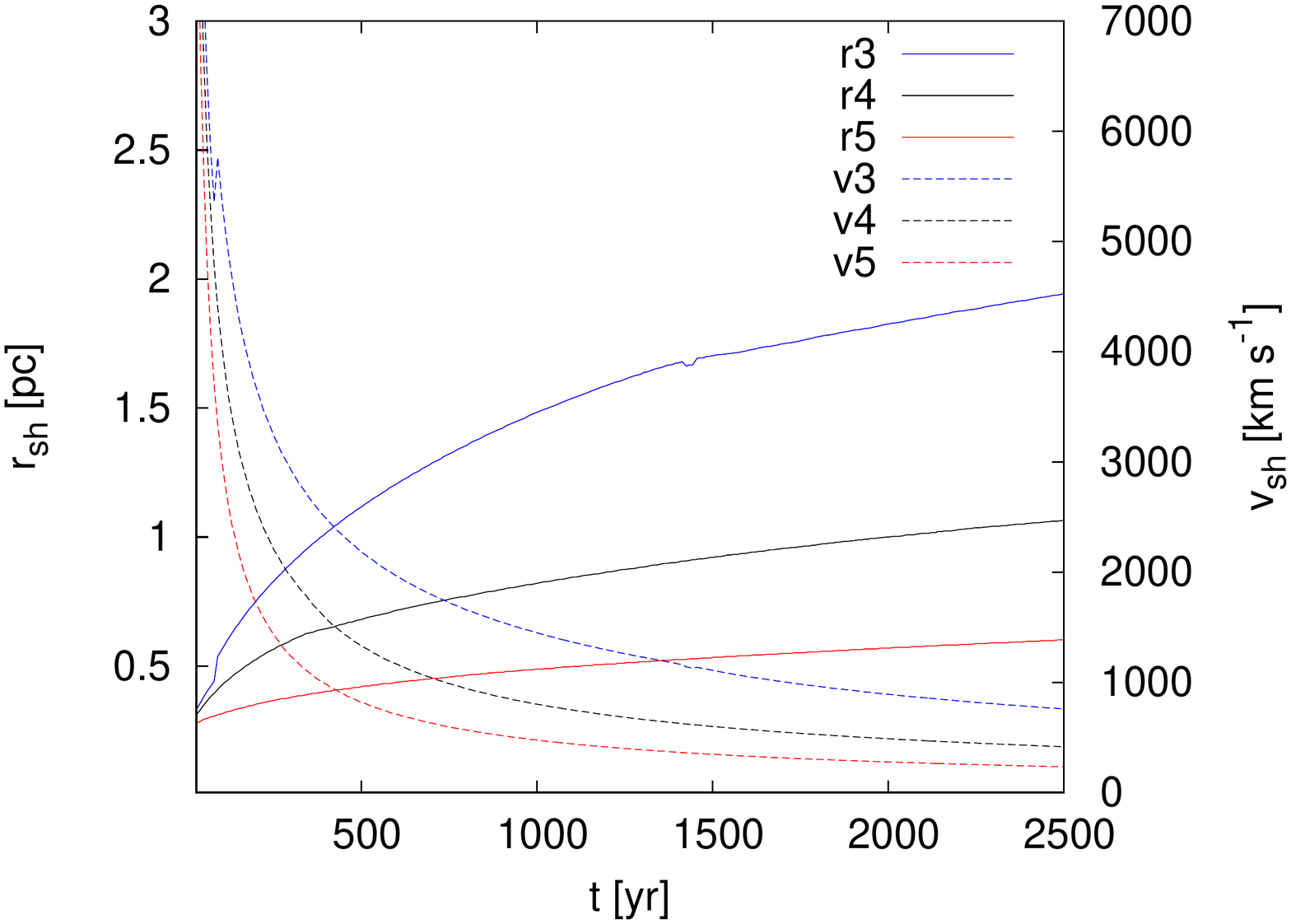}
\includegraphics[angle=0,width=0.49\linewidth]{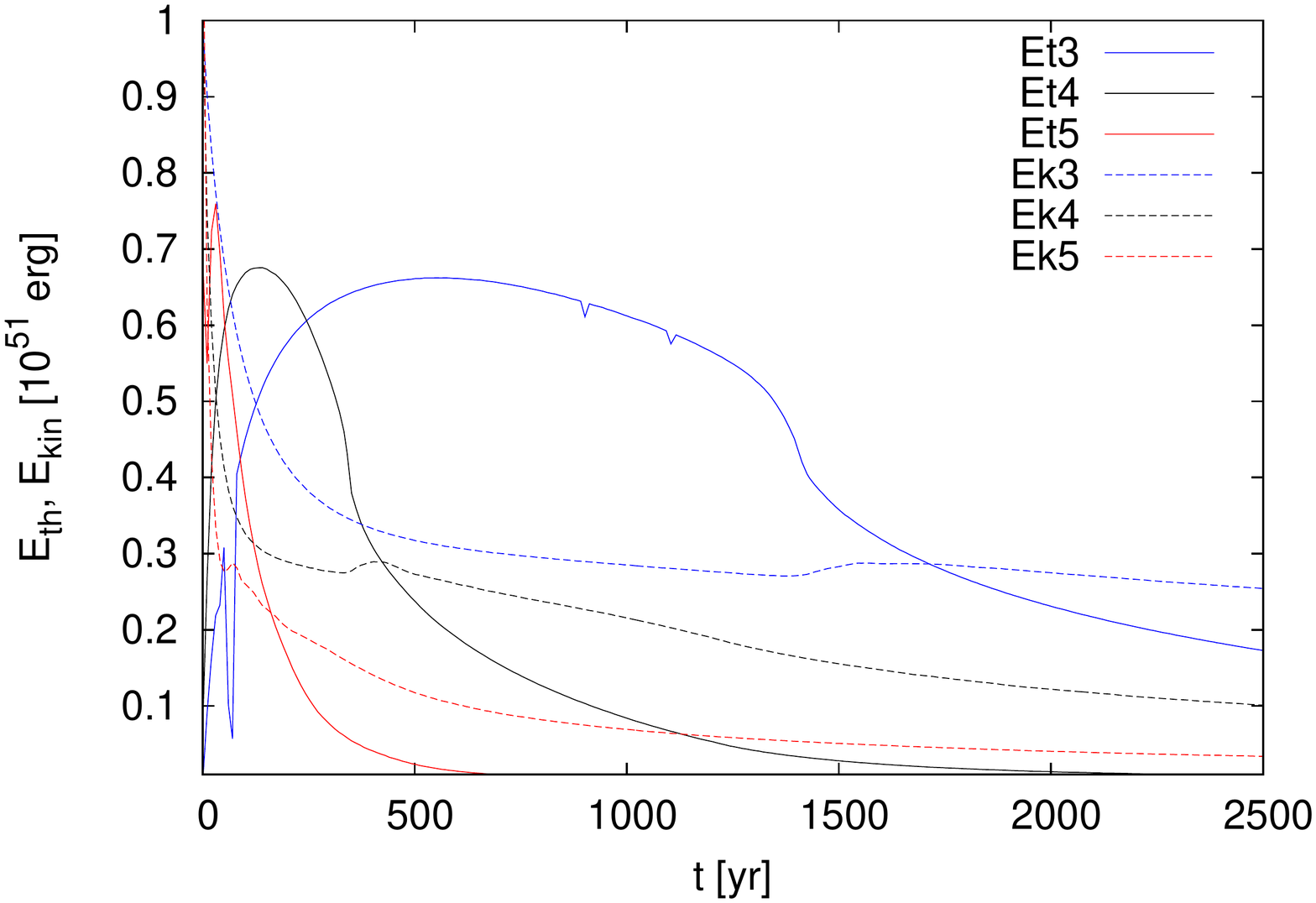}
\caption{One-dimensional FLASH simulations. Left panel: time evolution of the position of the forward shock, $r_{sh}$ (r3, r4, r5), and of its expansion velocity, $v_{sh}$ (v3, v4, v5). Right panel: Time evolution of the thermal energy inside the shock, $E_{th}$ (Et3, Et4, Et5), and of the kinetic energy of the shock, $E_{kin}$ (Ek3, Ek4, Ek5). Individual quantities are labeled with the value of the power index of the ISM density $n_{out} = 10^3, 10^4$, and $10^5$ cm$^3$.} 
\label{time-evolution}
\end{figure*}

\begin{figure*}
\centering 
\includegraphics[angle=0,width=0.75\linewidth]{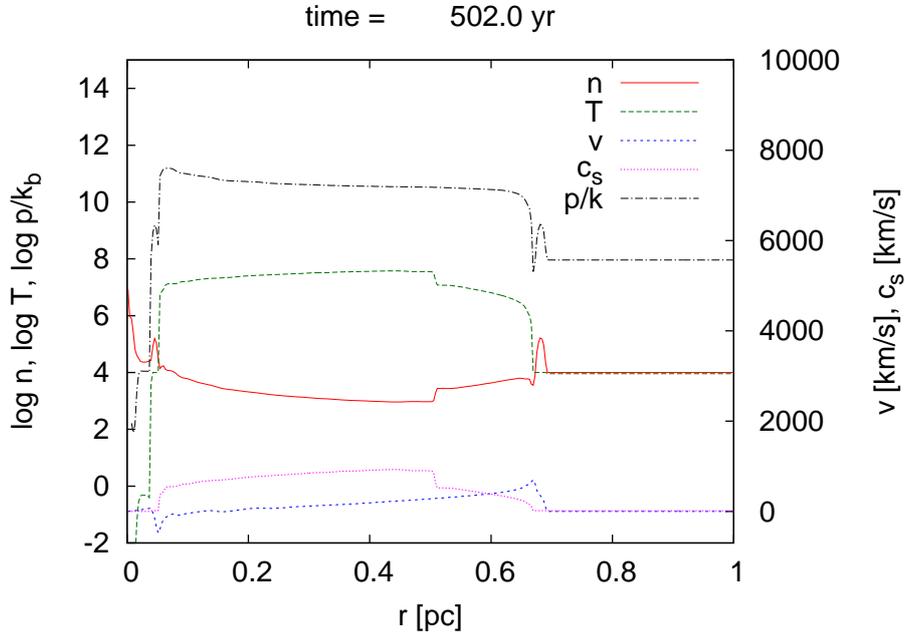}
\caption{One-dimensional FLASH simulations: profiles of the density $n$, the temperature $T$, the velocity $v$, the sound speed $c_s$ and the pressure $p/k_B$  for  $r < 1$ pc at time = 502 yrs in a medium of density $n_{out} = 10^4$ cm$^3$.} 
\label{flash-profile}
\end{figure*}

 The time evolution of the position of the forward shock, $r_{sh}$, and of its expansion velocity, $v_{sh}$, and the time evolution of the thermal energy inside of the shock, $E_{th}$, and of the kinetic energy of the shock, $E_{kin}$, are shown in Fig. \ref{time-evolution}.
After some time a thin shell forms behind the leading shock. At this thin shell formation time, $t_{sh}$, the thin shell has radius $r_{sh}$ and mass $m_{sh}$. The thermal energy inside the cavity of the shell, $E_{th}$, and the kinetic energy of the shell, $E_{kin}$,  at time $t_{sh}$ are given in Table \ref{table:1}. The rest of the initial injected energy, $E_{tot}$, was radiated away.
 At this stage of SN expansion, when the thin shell is formed, we switch to the code RING using the infinitesimally thin shell approximation.  
 In Fig. \ref{flash-profile} we show profiles of density, temperature, velocity, speed of sound, and pressure for $n_{out} = 10^4$ cm$^3$ at the time $t_{sh} = 502$ yr.

The 3D simulation with the code RING starts from the radius $r_{sh}$  at the time $t_{sh}$  given in Table \ref{table:1} when, according to the 1D FLASH simulation, the thin shell begins to form.
Formation of the thin shell is connected to the rapid loss of thermal energy due to the increased density that leads to enhanced cooling. The time of the shell formation was set on the basis of FLASH simulations and the choice was tested by a comparison of FLASH and RING results.
Into this initial volume we inserted at the time $t_{sh}$ the initial thermal energy  $E_{th}$.
The shell of total mass $m_{sh}$ (see Table \ref{table:1}) gets the initial expansion velocity corresponding to kinetic energy  $E_{kin}$ defined by FLASH simulations.
The initial velocities of individual elements have a constant value in all directions from the SN position, meaning that in all simulations of expanding shells we disregard the rotation of the ISM in setting the initial velocities of the shell elements. This is a valid assumption because the shell size at $t_{sh}$ is still small compared to the scale over which the local velocity field changes. 
 
 \begin{table*}
\caption{Formation of the thin shell and cooling time of the internal volume.}
\label{table:1}
\centering
\begin{tabular}{|r l r r r r r|}
  \hline\hline
 $n_{out}$ &  $t_{sh}$ &  $r_{sh}$ & $m_{sh}$ & $E_{th}/E_{tot}$ & $E_{kin}/E_{tot}$ & $t_{cool}$ \\
 $[cm^{-3}]$ & [yr] & [pc] & [$M_{\odot }$] & ~ & ~ & [kyr] \\
 \hline
  $10^3$ & 2021 & 1.83 &   403 &   0.23 &   0.28  & 46  \\
  $10^4$ & 502  & 0.68  &  210 &   0.24 &   0.27 & 9  \\
  $10^5$ &  150  & 0.33 &  177 &   0.23 &   0.21 & 3  \\
\hline
\end{tabular}
\end{table*}

\section{Expansion of the shell: Code RING}

The expansion of the infinitesimally thin shell in 3D is described with the code RING. The supersonically expanding shell accumulates the ambient medium and slows down. The code RING was originally written by Jan Palou\v s
(2D version: \citet{tenorio-tagle1987}; 3D version: \citet{1990IAUS..144P.101P}) and later extended by So\v na Ehlerov\' a
, most recently used in \citet{2018A&A...619A.101E}. The code RING is currently not publicly available, and a more detailed description will be given on the web page <http://galaxy.asu.cas.cz/pages/ring>.  There have been other codes that use the Kompaneets approximation (\citet{silich1996}; \citet{1999A&A...343..352M}); the main difference between these and RING lies in the treatment of conditions inside the bubble. 

The infinitesimally thin shell is divided into $N_{elem}$ elements.
They are originally distributed equally across the initial sphere;
we use the HEALPIX \citep{2005ApJ...622..759G} distribution with 192 elements.
The subsequent
evolution may distort the shell so that the distribution of elements
does not remain homogeneous and their sizes (= areas) vary.

For each of the elements we solve the momentum and mass conservation equations. The momentum 
equation is 
\begin{eqnarray}  
  {d \over dt}(m_{\mathrm{i}} \vec{v_{\mathrm{i}}}) 
  + \rho_{\mathrm{out}}  
  [{\vec{S_{\mathrm{i}}} \cdot (\vec{v_{\mathrm{i}}}} - \vec{v_{\mathrm{out}}})] \vec{v_{\mathrm{out}}} 
  = (P_{\mathrm{int}} - P_{\mathrm{out}})\vec{S_{\mathrm{i}}}  + m_{\mathrm{i}}\vec{g}, 
  \label{momentum}  
  \end{eqnarray}
  where $m_{\mathrm{i}}$ is the mass of the i-th element,  
      $\vec{S_{\mathrm{i}}}$ is its surface area (i.e., the surface area multiplied by the normal vector),   
      $\vec{v_{\mathrm{i}}}$ is its velocity of expansion,  
      $\vec{v_{\mathrm{out}}}$ is the velocity of the medium outside of it,  
      $P_{\mathrm{out}}$ and   
      $\rho_{\mathrm{out}}$ are the pressure and density of the medium outside of the shell,  
$\vec{g}$ is the gravitational acceleration of the NSC and SMBH 
, and $t$ is the time.

The shell sweeps up the ambient medium as it expands: 
for each surface element, when the velocity component perpendicular to the element, 
$v_{\perp} = (\vec{v_{\mathrm{i}}} - \vec{v_{\mathrm{out}}})_{\perp}$, exceeds the local speed of sound, the mass accumulation is given as  
\begin{equation}  
  {dm_{\mathrm{i}} \over dt} = 
  \rho_{\mathrm{out}} \ [(\vec{v_{\mathrm{i}}} - \vec{v_{\mathrm{out}}}) \cdot \vec{S_{\mathrm{i}}}]. 
  \label{mass}  
\end{equation}

When $v_{\perp}$ drops below the sound speed in the ISM, the accumulation of mass is stopped and the element continues its expansion in the gravitational potential in a ballistic way without accumulating mass further.

The infinitesimally thin shell approximation carries with it the assumption that the shell has such a high density
that it is able to radiate away all the thermal energy produced by compression of the ambient medium by the shock.
The pressure inside the cavity of the shell $P_{\mathrm{int}}$ follows the equation    
\begin{equation}  
  P_{\mathrm{int}} = {2 \over 3} {E_{th} \over V_{\mathrm{int}}},
\end{equation}
where $V_{\mathrm{int}}$ is the total volume inside the shell.

We disregard the radiative cooling of the hot and low density medium inside the cavity of the shell. Its  thermal energy, $E_{th}$,  follows the adiabatic    
energy balance equation
\begin{equation}  
  {dE_{th} \over dt} =
  - {dV_{\mathrm{int}} \over dt} P_{\mathrm{int}}.  
\label{energy}  
\end{equation}

Integrations are done using the Runge-Kutta method (a fourth order, RK4) with
  an adaptive step size. The surface areas of expanding elements are
computed at every time step by interpolation between the positions of neighbors.

To check the importance of radiative cooling  from the interior cavity of the shell, we follow the shell expansion with the 1D code FLASH, and we record the time $t_{cool}$ when the medium inside the shell, originally heated by the reverse shock to high temperatures, cools down to $10^4$ K (see Table \ref{table:1}). For $t > t_{cool}$ the internal pressure $P_{int}$ drops sufficiently to be ignored, and the shell enters the momentum-driven snowplow phase. We examine the results of simulations with the code RING when the internal pressure is switched off at $t_{cool}$.  In this case the RING simulations give the total mass collected in the shell as being only $\sim 1\%$ lower compared to the cases when the internal pressure acts all the time. This demonstrates that the importance of cooling from the interior of the shell is small: the shell expansion is mainly governed  by momentum with little  contribution from the internal pressure.

\begin{figure*} 
\centering 
\includegraphics[angle=0,width=0.45\linewidth]{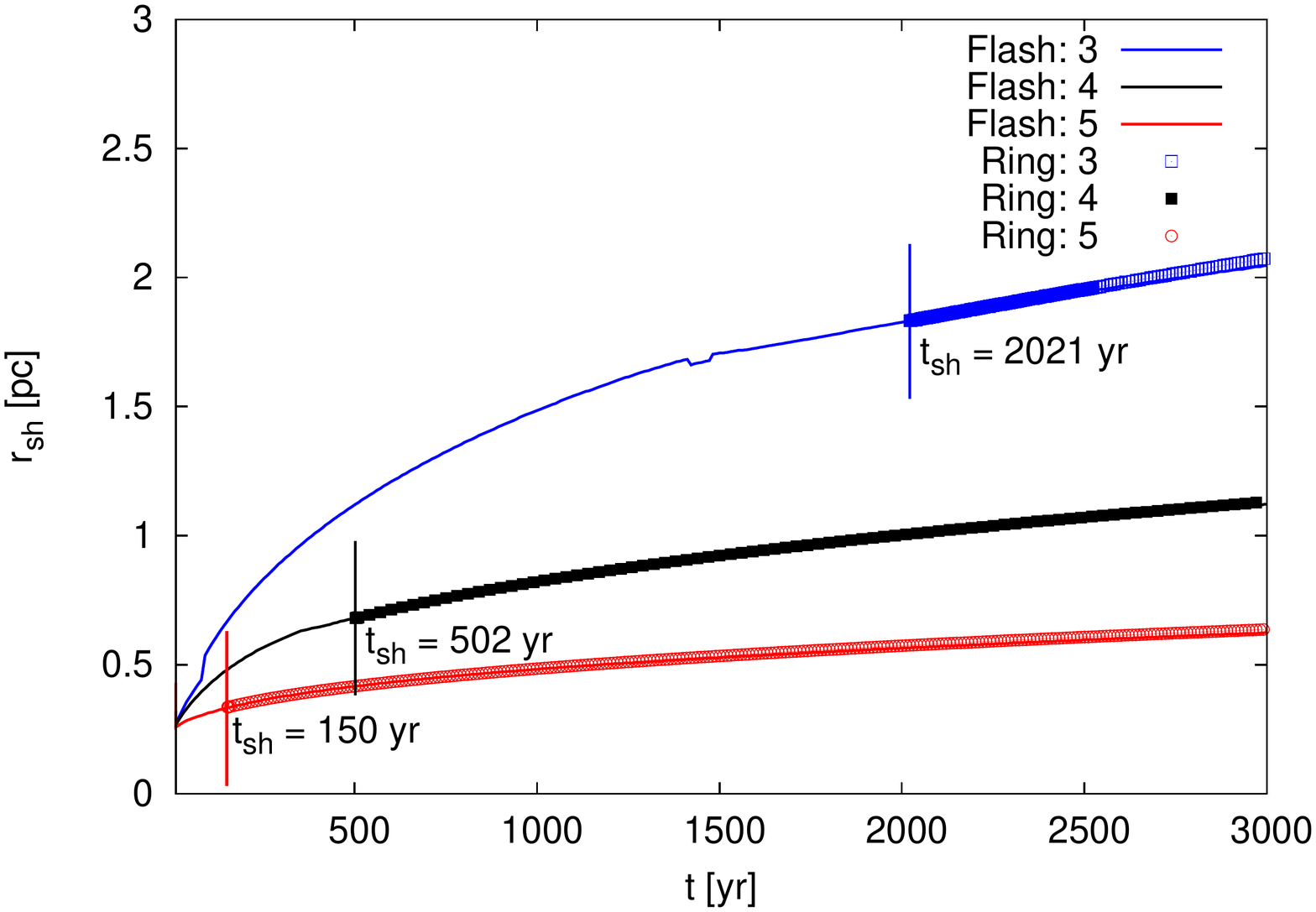}
 \includegraphics[angle=0,width=0.45\linewidth]{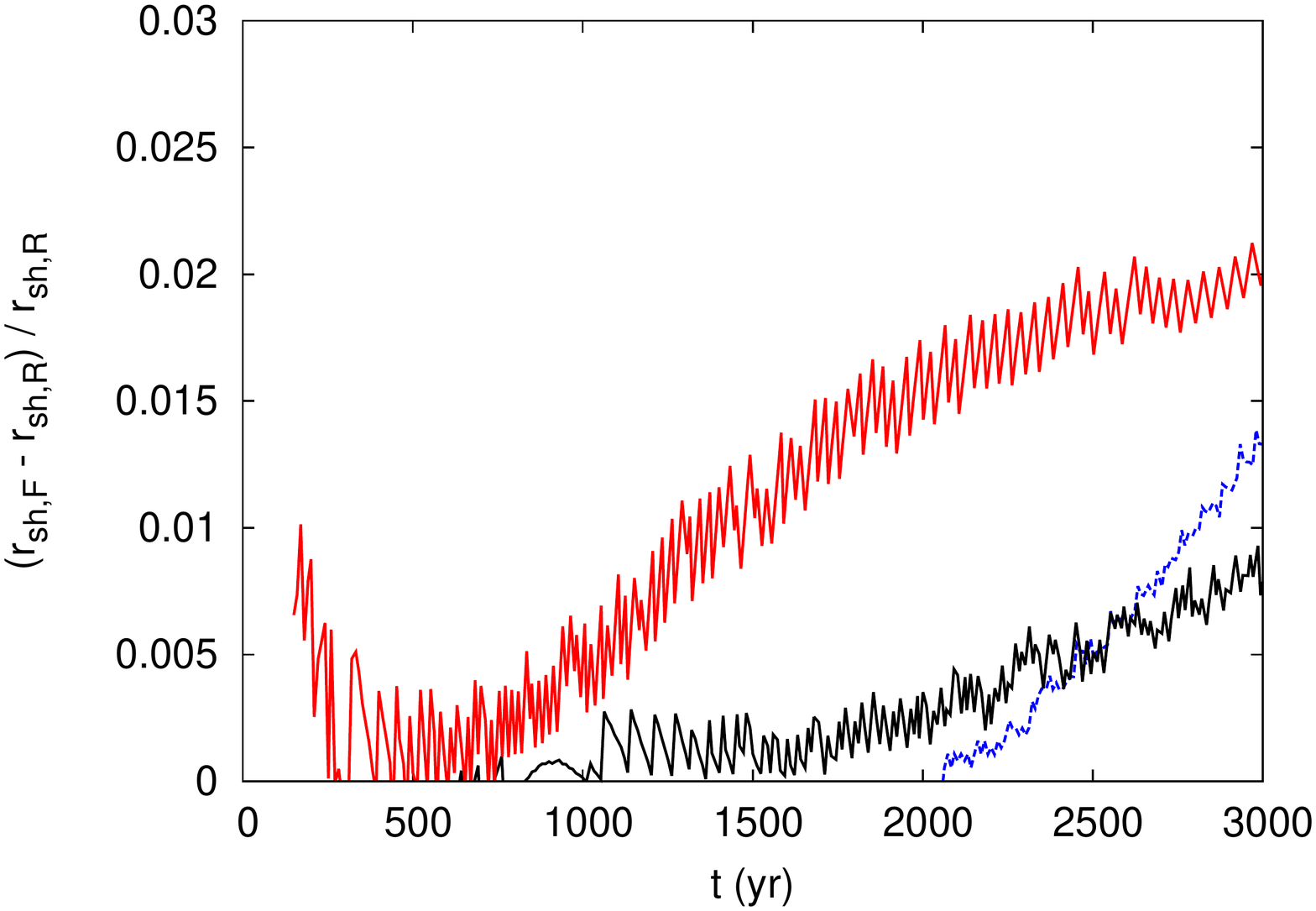}
\caption{Shell radius, $r_{sh}$, as a function of time during the first 3000 years of expansion. The 1D FLASH is joined by the 3D RING at $t_{sh}$: absolute values (left panel), relative values of the difference $(r_{sh,F} - r_{sh,R})/r_{sh,F}$ (right panel), where $r_{sh,F}$ is the radius of the shell according to FLASH and $r_{sh,R}$  is the radius of the shell according to RING simulations. The number in the line label shows the power index of the value $n_{out} = 10^3, 10^4$, and $10^5$ cm$^{-3}$.} 
\label{flash-ring}
\end{figure*} 

As an additional test of the validity of the transition from FLASH to RING, in the static medium without the external gravity of the NSC and SMBH, we computed the shell expansion during the first 3000 years. Figure \ref{flash-ring} compares the radius of the shock as a function of time from  FLASH and RING. As is visible in Fig. \ref{flash-ring},
there is a close correspondence in shell radii: the difference at 3000 yr is still below 2\% demonstrating a very good agreement achieved between these two very different simulations.

\section{Later evolution: Simulations with the code RING}

The evolution of the expanding thin shell after its formation at $t_{sh}$ is followed with the code RING. To explore expansions from different positions relative to the SMBH, we changed the initial galactocentric distance, $r_{GC}$, and the elevation angle, $\theta_{GC}$, which is the altitude of the SN explosion above the galactic plane as seen from the galactic center.
We computed the total mass collected in the shell, $m_{sh, tot}$, at the end of mass accumulation, when all its elements slow down and expand with subsonic velocity.

\begin{figure*} 
  \centering
\includegraphics[angle=0,width=1.1\linewidth]{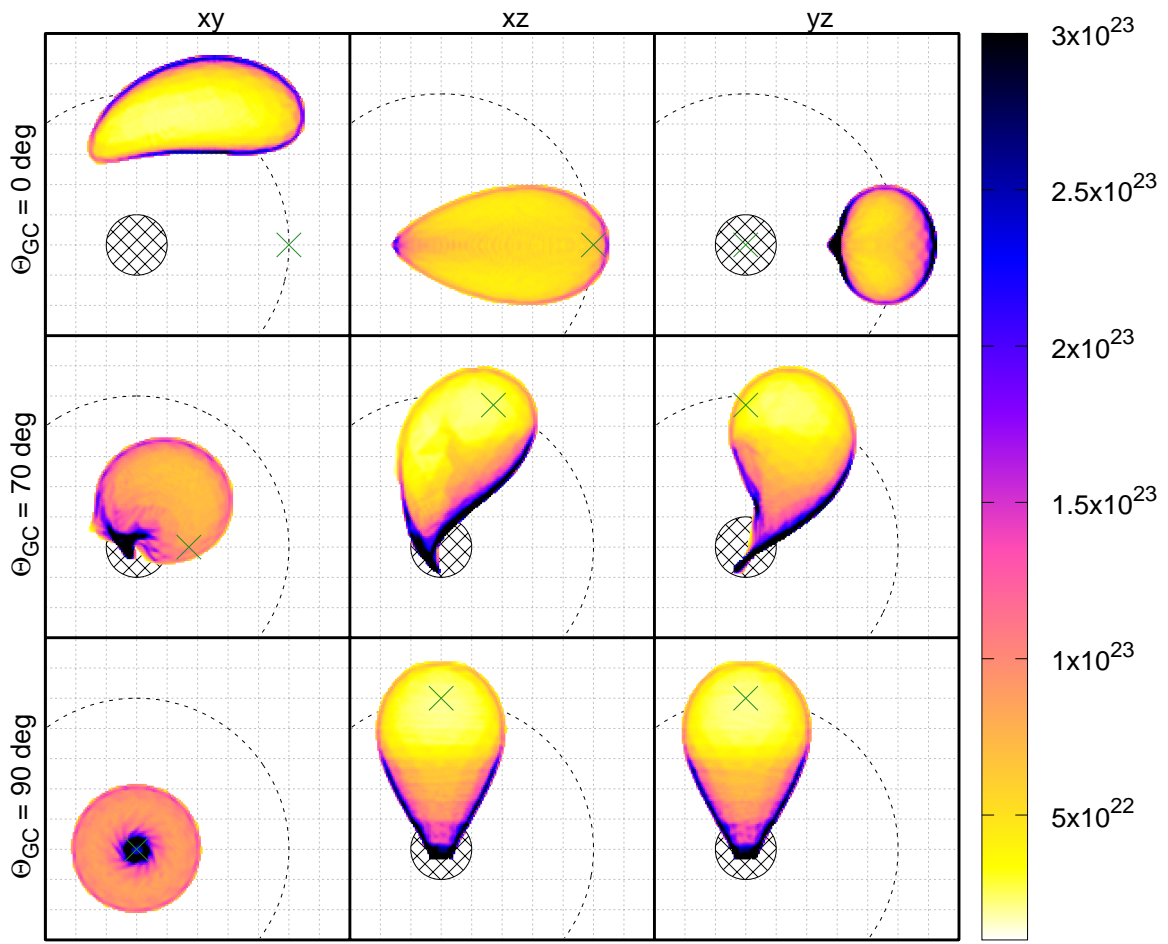}
\caption{Three-dimensional simulations with the code RING. We show the projected shell column density at 60 kyr after the formation of the shell. The SN explodes at $r_{GC} = 5$ pc and $\theta_{GC} = 0^\circ$ (top), $70^\circ$ (middle), $90^\circ$ (bottom) in an ambient medium of density $n_{out} = 10^4$ cm$^{-3}$, with $M_{SMBH} = 4 \times 10^6 M_\odot$ and the NSC as given in Sect. 2. The color bar gives the projected density in units of cm$^{-2}$. The meshed region shows the central region with a radius of 1 pc. The dashed line shows the reference circle with a radius of 5 pc and the cross gives the original position of the SN explosion.
} 
\label{ring-rgc5}
\end{figure*}

As an example, in Fig. \ref{ring-rgc5} we show the projected column density of the 3D shell along three viewing axes in galactocentric Cartesian coordinates (x, y, z)  at a time of 58 kyr after the formation of the thin shell. The size of the box is 20 x 20 x 20 pc and the thin shell expands into the homogeneous ambient medium of density $n_{out} = 10^4$ cm$^{-3}$ from SN explosions at $r_{GC} = 5$ pc, $\theta_{GC} = 0^\circ, 70^\circ, 90^\circ$.

As shown in Fig. \ref{ring-rgc5} (top panel), at $\theta_{GC} = 0^\circ$, where the galactocentric gravitational attraction is balanced with the centrifugal force, the expanding shell is deformed by differential rotation, creating elongated shapes when projected onto the galactic plane, wrapping around the Galactic center. In this case, at $200$ kyr, when all the shell elements become subsonic, $m_{sh, tot} = 6 300$ M$_{\odot}$.

Another deformation is due to the gravitational attraction, $F_{\rm z}$, parallel to the rotational axis, which is an important part of the total gravitational force not balanced by the rotation, stretching the shell toward the symmetry plane of the galaxy.
For $\theta_{GC} = 70^\circ$ (Fig. \ref{ring-rgc5}, middle panel), the shell elements follow helical orbits with parts of the shell spiraling toward the center of gravity.
In this case, its total mass is $m_{sh, tot} = 15 000$ M$_{\odot}$.

In the case in which $\theta_{GC} = 90^\circ$ (Fig. \ref{ring-rgc5}, bottom panel), when the SN explodes on the rotational axis, there is almost no rotation and the expansion is influenced by the galactocentric gravitational force only. Its total mass is  $m_{sh, tot} = 10 500$ M$_{\odot}$.

The mass of the shell corresponds to the volume originally occupied by the ISM that was collected in the shell elements. It is influenced by the length of the path on which an element has moved supersonically relative to the motion of the ISM. The longest supersonic path is at middle elevations around $\theta_{GC} \sim 70^\circ$, where the rotation combines with the motion towards the galactic plane (see the next section for more detailed analysis).

\section{Results}

\subsection{Angular momentum of the swept up gas}

We explore the distribution of the angular momentum, $L(R, z)_i$, of shell elements at the time when their expansion velocity drops below the speed of sound. At this time, the mass accumulation is stopped and the mass $m_i$ of an element and its angular momentum $L_i$ are fixed. The angular momentum, $L(R, z)_i$, of an individual shell element defines the ``angular momentum radius'', $R_L$, which is the distance from the rotational axis where a body at the circular orbit at $z = 0$ has the same specific angular momentum (angular momentum per unit mass):
\begin{equation} 
R_L \ V_{rot}(R_L, z = 0) = L(R, z)_i / m_i.
\label{Ram}
\end{equation} 
The distribution of mass as a function of angular momentum radius $R_L$ represents the mass distribution in a flat disk, where the two components of velocity, the component away from the rotational axis and the component parallel to the rotational axis, are suppressed. The only remaining component is the circular rotational component. Actually, $R_L$ reflects the centrifugal barrier of the matter collected in expanding SN shells.

\begin{figure*} 
\centering 
\includegraphics[angle=0,width=0.77\linewidth]{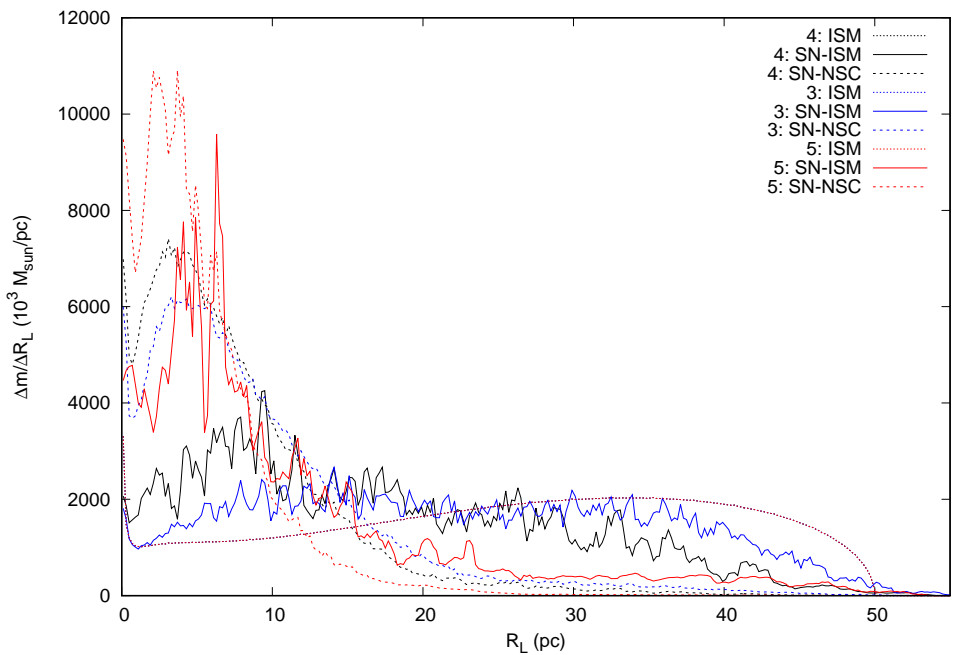}
\caption{Distribution of mass as a function of $R_L$. We show a fraction of the total mass $\Delta m$ in the $\Delta R_L$ bins as a function of $R_L$ for three values of $n_{out}$, with $M_{SMBH} = 4 \times 10^6 M_\odot$ and the NSC as given in Sect. 2. The number in the line label shows the power index of the value $n_{out} = 10^3, 10^4$, and $10^5$ cm$^{-3}$. The vertical scale holds for $n_{out} = 10^4$cm$^{-3}$; for a lower or higher ISM density it should be changed to ten times lower or higher values, respectively. 
} 
\label{results-am}
\end{figure*}

We compare the distribution of mass as a function of $R_L$: in Fig. \ref{results-am} we plot the fraction of mass $\Delta m$ distributed in $\Delta R_L$ bins as a function of $R_L$ in the following cases: 
\begin{enumerate}
\item ISM: A homogeneous ISM with rotation given by Eq. (\ref{eq-rotation})
  at galactocentric distances $r_{GC} < 50$ pc.
\item SN-ISM:
  The total mass of elements of expanding shells at the time of crossing the sound speed
  from SNe distributed randomly within the computational region following a uniform density distribution. The total mass and angular momentum depend on the total number of SNe included in the computation. Here, the total number of SNe is adjusted so that the total mass in all expanding SNe elements (at the time when they stop mass accumulation) is the same as the mass of the homogeneous medium inside the computational region that was used in the previous case, taking into account the density of the ISM. 
\item SN-NSC: The total mass of elements of expanding shells at the time of crossing the sound speed
from SNe distributed with a radial profile corresponding to the density distribution in the young NSC$_1$ as described by Eq. (\ref{NSC-density}). The total number of SNe is the same as in the case SN-ISM.    
\end{enumerate}

These three cases are compared in Fig. \ref{results-am} where we show the mass, $\Delta m$, in $\Delta R_L$ bins  as  a function of  $R_L$. 
The mass distribution as a function of $R_L$ for the homogeneous ISM shows a narrow central peak composed of the matter residing along the rotational axis with $L_i = 0,$ and some more ISM elements with small $L_i$, since
the $L_i$ of elements of the rotating ISM at a given $R$ decreases with $z$ proportional to  $R/\sqrt{R^2 + z^2}$. Consequently the region of small $L_i < R_L V_{rot}(R_L, z = 0)$  has a conical shape around the rotational axis. This region, where the mass with a small angular momentum resides, is separated from the regions with a mass of higher angular momentum by a separation line at 
\begin{equation}
{{V_{rot}(R, z)} \  R \over V_{rot}(R_L, z = 0)} =  R_L.
\end{equation}

The homogeneous ISM shows an increasing profile of ${\Delta m \over \Delta R_L}(R_L)$ (see Fig. \ref{results-am}); its slope reflects the shape of the rotation curve as the rotation slows down with distance $z$ off the symmetry plane.
The distribution has a maximum near $\sim 35$ pc and it decreases for $R > 40$ pc, which is due to the finite size of the computational region (in our case 50 pc from the galactic center). The three densities of the ambient medium are compared in Fig. \ref{results-am}. In the case of the homogeneous ISM distribution, they show identical profiles (providing they are normalized to the same total mass).

In the second case, with the homogeneous distribution of SNe within the
computational region, for $n_{out} = 10^4$ cm$^{-3}$ there is an increase up to the maximum at $\sim 10$ pc followed by a decrease for $R_L > 10$ pc (see Fig. \ref{results-am}). With $n_{out} = 10^5$ cm$^{-3}$, the maximum is at a distance of $\sim 7 $ pc and it has a higher value, and with $n_{out} = 10^3$ cm$^{-3}$ the profile is much flatter, with a maximum at $\sim 17$ pc at the lower maximum value.

Compared to the previous case of a homogeneous ISM distribution, some of the
matter with $R_L > 25$ pc is shifted to lower $R_L$ values forming the
maximum at lower values of $R_L$.
This is due to the fact that the motion of those elements of the expanding SNe
shells at high $z$ distances from the galactic plane with low angular momentum is accelerated by the part of the galactocentric gravitational attraction of the 
SMBH and NSC not balanced by rotation. This force, which is 
perpendicular the galactic plane, is the reason why those elements moving
toward the galactic plane collect more mass than they would collect without
this perpendicular force. The motion of elements at low $z$ distances with
high angular momentum is not influenced by this perpendicular attraction (or
less compared to the high $z$ cases) and the mass and momentum accumulation
is stopped earlier.

In the third case, the total number of SNe is the same as in the case of the homogeneous distribution. As one may see in Fig. \ref{results-am}, there is less mass particularly in the outer parts at $R_L > 5$ pc, and the maxima shift below  $R_L \sim 5$ pc, close to the size of the core of the density profile of the young NSC$_1$, where most of the SNe explode.

\subsection{Feeding the central region}

We derive how much of the shell mass is delivered into the central
region, defined as a sphere with radius,
$r_\mathrm{feed}$, which is a free parameter in this model.
The physical motivation behind it is that SMBHs are often
surrounded by gaseous circumnuclear disks, and any gas
moving through the volume occupied by such a disk will probably
impact it and merge with it in dissipative shocks and then
eventually will be accreted onto the SMBH. In the Milky Way, the
radius of the circumnuclear disk is on the  order of several
parsecs \citep{2018PASJ...70...85T}. Here
we stay conservative and set $r_\mathrm{feed} = 1$\;pc.

\begin{figure*} 
\centering 
\includegraphics[angle=0,width=0.77\linewidth]{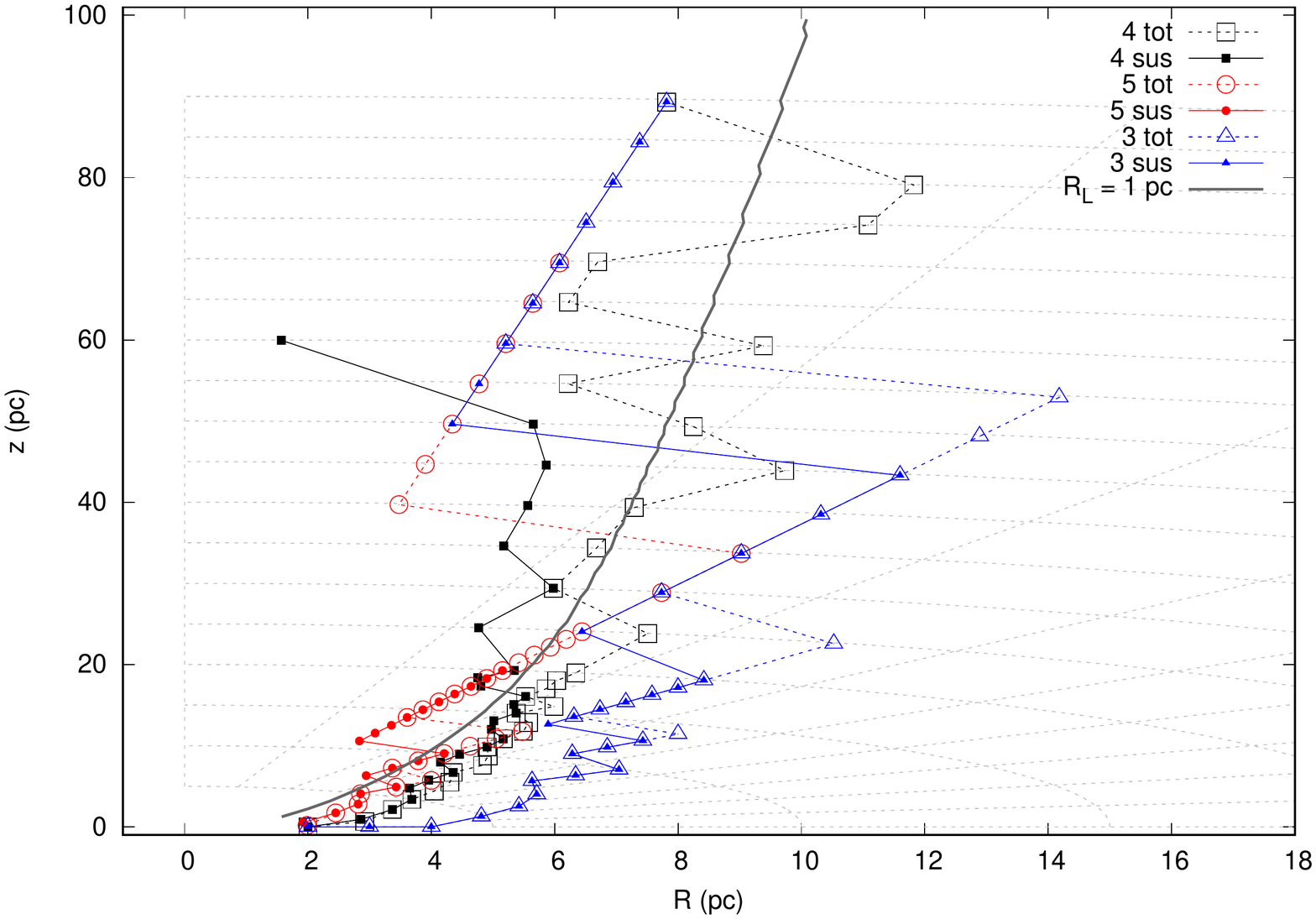}
\caption{Border lines for feeding the central 1 pc in the $(R,z)$ plane. Black continuous line: $R_L = 1 pc$ line separating the low and high angular momentum places. Colors represent different values of  $n_{out}$ in particles per cm$^{-3}$: blue - $10^3$; black - $10^4$; red - $10^5$, with $M_{SMBH} = 4 \times 10^6 M_\odot$ and the NSC as given in Sect. 2. The open symbols show feeding of the central $1 pc$ by SNe shell elements, solid symbols show feeding by supersonic elements only. The thin dashed lines give places of constant $r_{GC}$ (in 5 pc steps) and $\theta_{GC}$ (in 10$^\circ  $ steps).
} 
\label{divline}
\end{figure*}

The  mass feeding depends on the position of the SN explosion.
Figure \ref{divline} shows the line in the $(R, z)$ plane where
\begin{equation}
{{V_{rot}(R, z)} \  R \over V_{rot}(1, 0)} = 1 pc
\end{equation}
(black continuous line) separating the places of low angular momentum (to the left of it) from high angular momentum places (to the right from it). It is compared to 
lines based on RING simulations separating the places of the SNe explosions from where the shell elements reach the central 1 pc (to the left of the  zigzag curve), and places from where the shell elements do not reach the central 1 pc (to the right of the zigzag curve). We show the results for different values of $n_{out}$ and we distinguish the mass delivered by supersonic elements (solid lines) from mass that is delivered to the central 1 pc by all elements (dashed lines).

The zigzag lines in Fig. \ref{divline} generally follow the solid line separating the low and high angular momentum places, but some feeding of the central $1 pc$ can also be achieved even from high angular momentum places. This is  more likely close to the galactic center at $r_{GC} < 20$ pc for expansions in the media having lower values of $n_{out}$. It can be explained as being due to the larger size of shells expanding in media with a lower density compared to shells expanding in media with a higher density. At larger distances from the galactic center at $r_{GC} > 20$ pc, the size of the shells is small compared to the galactocentric distance for any of the tested densities and the delivery to the central parsec follows the line separating low and high angular momentum places.

There is another result visible in  Fig. \ref{divline}. The mass feeding into the central 1 pc with supersonically expanding shell elements is restricted by the initial galactocentric distance,  $r_{GC}$: it happens from SNe inside of a maximum $r_{GC,feed,max}$, which  depends on the density of the ambient medium, $n_{out}$. It is 20, 50, or 140 pc for $n_{out} = 10^5, 10^4$ , or $10^3$ cm $^{-3}$; for larger distances no supersonic elements reach the central 1 pc.

\begin{figure*} 
\centering 
\includegraphics[angle=0,width=0.45\linewidth]{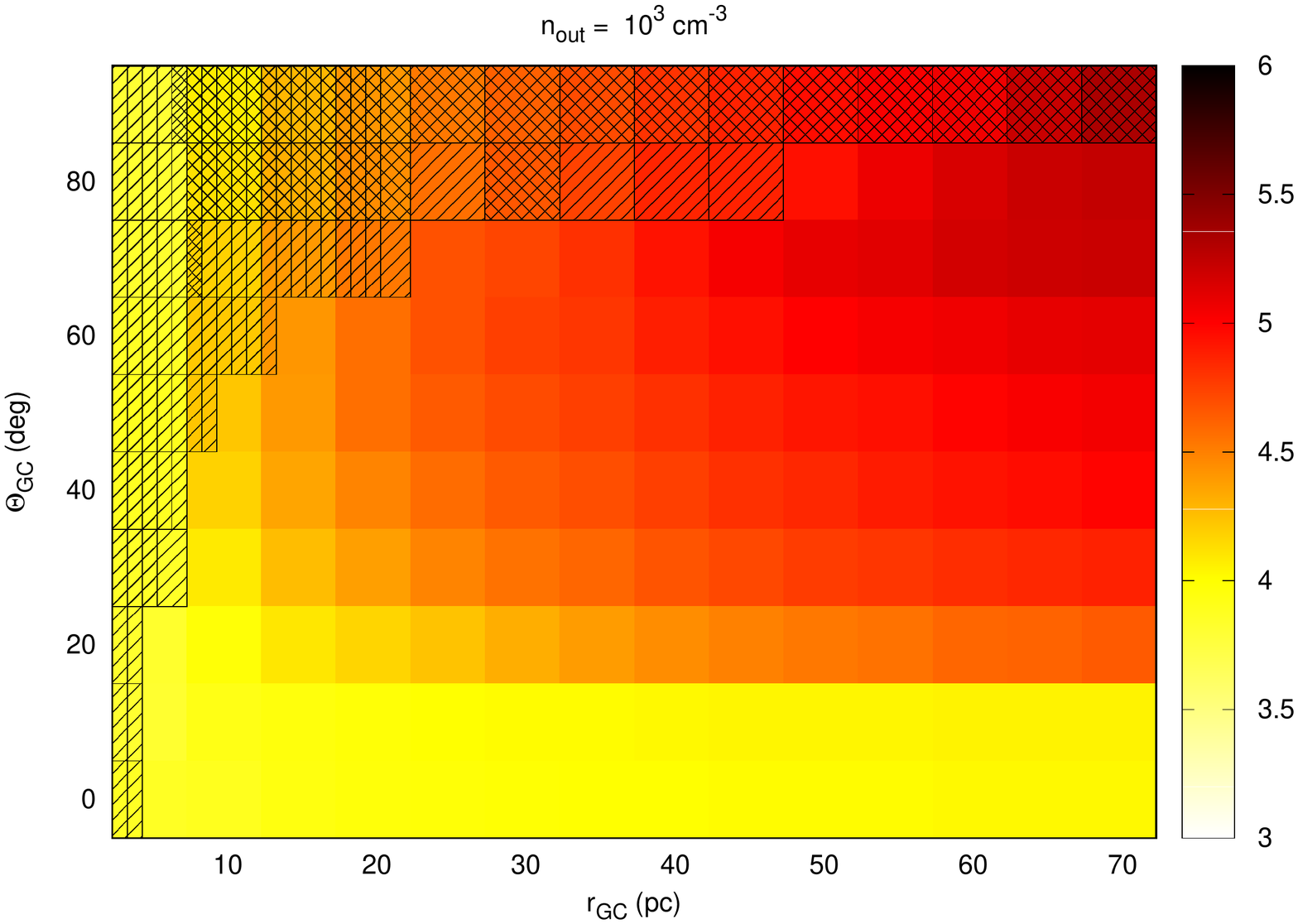}
\includegraphics[angle=0,width=0.45\linewidth]{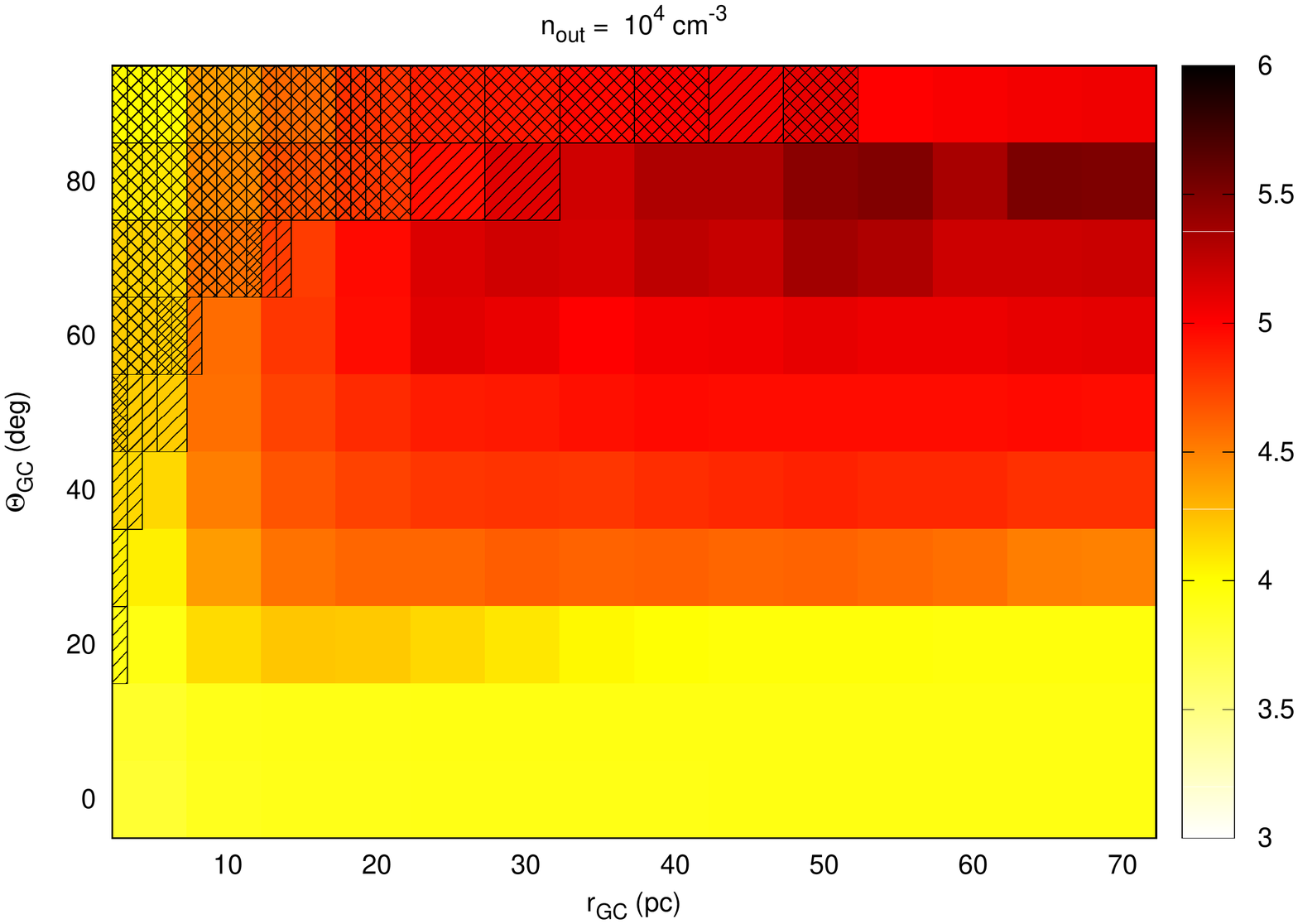}
\includegraphics[angle=0,width=0.45\linewidth]{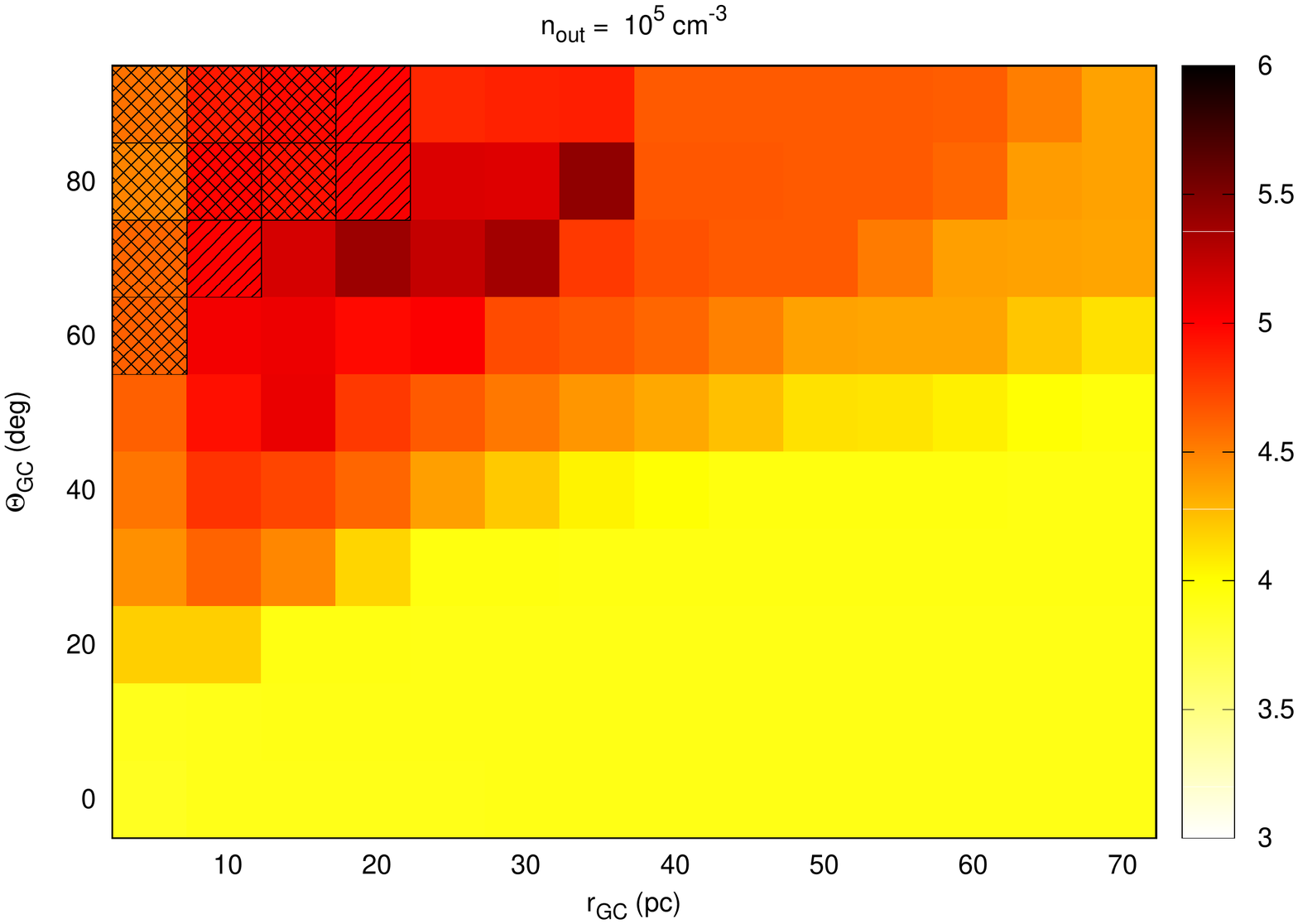}
\caption{The total mass collected in a single expanding shell at the end of mass accumulation, $m_{sh, tot}$,
  as a function of the galactocentric distance $r_{GC}$ and the elevation angle $\theta_{GC}$ for $n_{out} = 10^3, 10^4$, and $10^5$ cm$^{-3}$, with $M_{SMBH} = 4 \times 10^6 M_\odot$ and the NSC as given in Sect. 2. The mass delivery to the central parsec is shown with hatching: the areas without hatching do not deliver any mass, single hatched areas deliver up to 10\% of $m_{sh, tot}$, the double-hatched areas deliver more than 10\% of $m_{sh, tot}$.
} 
\label{results-2}
\end{figure*}

We examine the mass $m_{sh, tot}$, which is the total mass collected in a single expanding SN shell at the end of mass accumulation when all elements move with subsonic velocity. In Fig. \ref{results-2}, we show  $m_{sh, tot}$  at different ($r_{GC}, \ \theta_{GC}$) positions together with the fraction that is delivered by the supersonically moving elements of a shell into the central 1 pc. We show the results for three different values of $n_{out}$. The largest values of $m_{sh, tot}$ are achieved at certain values of $\theta_{GC}$, which grow with decreasing $n_{out}$ from $70^\circ$ to $90^\circ$.  The maximum value of $m_{sh, tot}$ increases from $10^5 M_\odot$ for $n_{out} = 10^5$ cm$^{-3}$ to $10^6 M_\odot$ for $n_{out} = 10^3$ cm$^{-3}$, and its position shifts away from the galactic center for lower values of $n_{out}$.


As can be seen in Fig.  \ref{results-2}, there is a delivery region from where the supersonic elements reach the central 1 pc.
We compute the average mass fed per SN, $\overline{m_{feed}^{100}}$, which is the average mass transported by one SN explosion taking place in the region $r_{GC} \le 100$ pc to the central region $r_{GC} \le r_{feed}$ = 1 pc. SNe are positioned randomly in  the region $r_{GC} \le 100$ pc with a specified  radial distribution (homogeneous, NSC$_1$).
  The product of the SN rate within the central 100 pc and the average mass fed per supernova, $\overline{m_{feed}^{100}}$, gives the mass flux $\dot M_{feed}^{100}$, which is the amount of mass transported per unit of time from the region within 100 pc into the central 1 pc via expanding SN shells.

  
The values of $\overline{m_{feed}^{100}}$ are given in Table \ref{table:2}. There  we also give
the fraction of the volume inside 100 pc occupied by an average expanding SN shell in supersonic expansion, $\overline{f_{SN}^{100}}$, the fraction of volume inside 100 pc occupied  by the delivery region, $f_{delivery}^{100}$, and an average time interval during which the shell is at least partly supersonic, $\overline{t_{sonic}^{100}}$. The last quantity multiplied by the SN rate within the central 100 pc gives the average number of SNe that should be visible within the central 100 pc.



\begin{table*}
  \caption{Type of  SN distribution. $n_{out}$ : the density of the ambient medium; $\overline{m_{feed}^{100}}$ : the average mass delivered into the central 1 pc region per SN exploding within the 100 pc from the SMBH; $\overline{f_{SN}^{100}}$ : the fraction of volume inside 100 pc occupied by an average expanding SN shell in supersonic expansion;
$f_{delivery}^{100}$ : the fraction of volume inside 100 pc taken by the delivery region; 
$\overline{t_{sonic}^{100}}$ : the average duration of SN shell.}
\label{table:2}
\centering
\begin{tabular}{|r r r r r r |}
  \hline\hline
 type & $n_{out}$ & $\overline{m_{feed}^{100}}$ &  $\overline{f_{SN}^{100}}$ & $f_{delivery}^{100}$ & $\overline{t_{sonic}^{100}}$  \\
 & [cm$^{-3}$] & [$M_\odot$] &  [$\times 10^{-5}$] & [$\times 10^{-3}$] & [kyr] \\
 \hline
SN-ISM & $10^3$ & 114 & 41.4 & 14.9 & 1070 \\
&        $10^4$ & 21 & 5.2 & 3.0 & 846 \\
&        $10^5$ & 9 & 0.1 & 0.68 & 668 \\
SN-NSC & $10^3$ & 205 & 28.5 & 14.9 & 465 \\
&        $10^4$ & 452 & 4.3 &  3.0 & 355 \\
&        $10^5$ & 1170 & 0.5 & 0.68 & 207 \\

\hline
\end{tabular}
\end{table*}

\section{Discussion}

Observations of Seyfert galaxies and AGNs \citep{2012ApJ...746..168D} show a link between the nuclear star formation rate, $\rm SFR$, and the black hole accretion rate, $\rm \dot M_{SMBH}$, that is approximated as  $\rm SFR \propto \dot M_{SMBH}^{0.8}$. AGNs at $z < 0.35$ exhibit an SFR correlated with X-ray luminosity, demonstrating the presence of AGN activity \citep{2020arXiv200711283Z}. The correlation between $\rm SFR$ and $\rm \dot M_{SMBH}$ was discussed by \citet[][and references therein]{2020arXiv200711285Z}: both quantities depend on the molecular gas mass within the central region, raising the question of how significant the above correlation is. However, the AGN activity evolves on much shorter timescales, $\rm <100 \ Myr$, than the star formation, $\rm >100\  Myr$, and rapid AGN variability occurs on even shorter timescales \citep{2014ApJ...782....9H}, thus  presenting a challenge for models connecting $\rm \dot M_{SMBH}$ to the $\rm SFR$.     

Here, we propose a connection between $\rm SFR$ and $\rm \dot M_{SMBH}$ based on expanding SN shells. The SN rate is certainly connected to the SFR: with a Salpeter initial mass function,
or with a top-heavy IMF, the SN rate is in any case proportional to the SFR. Within the central region, we do not separate the clumpy star formation in isolated molecular clouds from more distributed star formation inside the entire region: we assume that the star formation happens in the central region with a homogeneous or centrally condensed distribution similar to the scale of the $\rm NSC_1$.    

Observations of the Galactic center show the SFR = $3 \times 10^{-3}$M$_\odot$ /yr \citep{2004ApJ...601..319F,2004ApJ...604..662R,2013ApJ...771..118Z},
which corresponds to a SN rate of $3 \times 10^{-5}$ SN/yr. Similarly, a burst forming  $10^5 M_\odot$ of stars within 100 pc of the galactic
center will produce $\sim $ 1000  massive stars that will explode during the following 30 Myrs as SNe, which also gives an average SFR of  $3 \times 10^{-3}$M$_\odot$/yr, and a corresponding SN rate within the central 100 pc of 3 $\times 10^{-5}$/yr during those 30 Myr.
For homogeneously distributed star formation in an ISM of density $10^3 - 10^5$ cm$^{-3}$, the resulting supernovae would feed a total mass into the central 1 pc of $(9 \times 10^3 - 10^5)$ M$_\odot$ with an inflow rate of $\rm \dot M = (3 \times 10^{-4} - 4 \times 10^{-3})$ M$_\odot$/yr. If star formation were distributed in the same way as the $\rm NSC_1$ density distribution,
the total mass fed to the central parsec would be  $\rm (2 \times 10^5 - 10^6)$ M$_\odot$, with an inflow rate of $\rm \dot M = (7 \times 10^{-3} - 4 \times 10^{-2}) $ M$_\odot$/yr.

The SN rate within the central 100  pc - 3 $\times 10^{-5}$/yr  may be multiplied by the average duration of a SN shell as given in  Table \ref{table:2}, to give the number of expanding shells we would expect to see after a burst of star formation. This number varies between 6 and 35 for different ISM densities and for alternative choices of $NSC_1$ or homogeneous stellar distributions. The expanding SNe shells only fill a small fraction of the volume.

We show that an expanding SN shell can deliver considerable mass into the galactic center during epochs of star formation. The bursting AGN activity may be connected to mass delivery into the circumnuclear disk by individual expanding SN shells. Probably the SN shells feed the circumnuclear disk, which provides the mass to the central accretion disk. However, this accretion is limited by stellar winds and outflows due to feedback from the SMBH and young stars in the central star cluster. Thus, part of the inflowing mass may be pushed away, never reaching the nuclear SMBH.

However, AGNs show much higher values \citep{2020arXiv200711283Z}: $\rm SFR = 1 - 100$ M$_\odot$ /yr. With such a large nuclear SFR, the mass inflow via expanding shells may increase to $\rm 0.1 - 1000$ M$_\odot$/yr. This is just a crude estimate, since the shell versus shell interactions and supersonic shell collisions will influence the results and their importance should be evaluated.

The mass delivery by expanding SN shells also depends on the size of the circumnuclear disk. It can be only a fraction of a parsec as in NGC 4258 \citep{2018MNRAS.473.2198M} or it can be very large, $\sim 100$ pc, as in gas-rich AGNs \citep{2019A&A...623A..79C,2020arXiv200313280C}.
We explored the dependence of the mass delivery rate on the size of the delivery region briefly for $M_{SMBH} = 4 \times 10^6 M_\odot$: the total mass delivered scales with square of the radius of the circumnuclear disk, which probably reflects the geometry of the ``capture region''.  The mass delivery by expanding shells as a function of the size of the sphere of influence of the SMBH will be analyzed in a separate paper.

 A study of SN remnants in the vicinity of a SMBH was also described by
\citet{2015MNRAS.447.3096R}, with an emphasis on their X-ray properties. They adopt a similar method using the
Kompaneets (1960) approximation that assumes the wall of the
expanding SN shell is thin. This assumption is valid when the
thin expanding shell forms, which is connected to the rapid loss of
thermal energy due to the increased density leading to enhanced cooling.
\citet{2015MNRAS.447.3096R} use the thin shell approximation to describe also the earlier Sedov-Taylor phases when the expansion is still adiabatic, where the Kompaneets approximation may not be accurate. However, there are several differences compared with our treatment. In \citet{2015MNRAS.447.3096R}:
1) the ISM density distribution has a steep radial gradient,
in particular outside 1 pc from the galactic center;
2) the gravity includes the SMBH only, the NSC is not considered;
3) the distribution of massive stars and corresponding  SNe
is more concentrated in the galactic center, even when compared to the NSC$_1$ distribution adopted here.
Also, they do not account for the angular momentum of the ISM, which is critical to our treatment.
\citet{2015MNRAS.447.3096R} show SN remnant morphologies for different masses of the SMBH and for explosions at different distances from the SMBH. Here, we perform a more statistical study aimed at elucidating the inflow rates
as a function of ISM density, SNe radial distribution, and SNe location above the galactic plane.

Supernovae expanding shells with a low angular momentum that occur in a conical region near a rotational axis of a galaxy deliver mass to the central region, where it may settle into a circumnuclear disk.
When shell material passes within 1 pc of the central black hole, its orbital
time, 10$^5$ years $\times \left({10^6~M_{\odot}} \over {M_{BH}}\right)^{1/2}$,
approaches the expansion
time of the supernova shell, and the specific
angular momentum of the shell material will cause it to cease its inward motion
at the centrifugal barrier. 
At that point, the shearing shell will soon circularize, creating a
circumnuclear disk.

The SMBH is fed out of the disk, however its viscous time may be much greater compared to the disk feeding time by expanding SNe shells. Bursts of star formation invoke the scenario where the circumnuclear disk is formed quickly within a few tens of millions of years followed by longer periods of dissolution and continuous SMBH feeding out of the circumnuclear disk.

An
alternative idealized physical situation could be one in which a pressure
gradient supports the vertical stratification of the gas. However, there is
little evidence that the gas layer in galactic nuclei follows such a simple
description, except on relatively small scales (several parsecs) where
circumnuclear disks can be present. A more realistic description of the
interstellar medium would be one in which there are pronounced density
inhomogeneities and in which the vertical structure of the gas on $\sim 10-20$ pc
scales is determined by a stochastic velocity field caused by transient flows
and ejections caused by supernovae and winds from star formation complexes, by
accretion events, and by hydromagnetic activity. Our simplified treatment of
the interstellar medium is adopted in order to identify the important
dynamical implications of a supernova occurring in the strongly anisotropic
gravitational potential of a galactic nucleus. In later papers, we will
investigate how specific gas configurations such as circumnuclear disks or a
large-scale vertically stratified medium would affect the dynamics of a
supernova remnant. Ultimately, the consequences of a relatively strong,
ordered magnetic field will also be explored.

If the central parsec contains massive stars with strong winds, as our Galactic
center does, or if accretion
onto the SMBH leads to an outflowing wind, then the interaction of such winds
with an incoming supernova
shell will alter the dynamics, depending on the strength of the wind, but it would
not be likely to significantly
change the amount of ultimately accreted matter. This is because the momentum
of such winds is
predominantly radial, so it does not affect the specific angular momentum of the
material with which it
interacts, except possibly to reduce it \citep{2016MNRAS.459.1721B}. We have not
included such complications in our
models, but note that in specific cases such as the Galactic center, it can
embellish the evolutionary picture (see \citep{2005ApJ...635L.141R}).

Another complication that might be presented by a galactic nuclear region
is that, during the expansion time
of the models that we have considered, there is a fairly significant
chance that other SNe will occur
in the overall volume that we are considering (for example, a glance at
the radio image published by
\citet{2019Natur.573..235H} reveals a relatively high projected surface density
of SN remnants that
can be seen toward the Galactic center). So while we only consider the
dynamics induced by a single
SN, the more general case of multiple SNe might be occurring
a significant fraction of
the time, and could have consequences
such as strong shocks colliding with the shell, leading to some
redistribution of angular momentum that might favor mass flow into the central parsec, or reduce the mass delivery predicted by the isolated SN model.

If a supernova takes place inside the central parsec, it might not have time to
create much of a shell,
depending on the ambient density (c.f., Fig. \ref{flash-profile}), in which case the blast wave
would pass over the
SMBH, perhaps diverted to some extent by the SMBH magnetosphere or by any wind
produced
by the SMBH, but relatively little mass would be deposited there to be accreted.
Indeed, a SN
occurring within the central parsec could have the effect of partially or fully
clearing that region, depending
on how much interstellar gas is present there. The relevance of SN in the central
parsec is highlighted by
the fact that a massive young cluster with $\sim$100 O stars or massive
post-main-sequence stars
occupies the central half-parsec of our Galaxy, and furthermore, the overall
stellar density is maximized
in the central parsec, so that even Type Ia SNe would be concentrated
there if their progenitors
are present.
In the case of the Galactic center, the existing circumnuclear disk is too
massive to be expelled by a 
SN, but in any case, its inner edge has a $\sim $1 parsec radius, so
recent SNe might
be partially responsible for the relatively evacuated central cavity.

The very central part at $r_{GC} < 0.01$ pc will be strongly influenced by the activity of the SMBH.  The timescale for further dynamical evolution of the material inside $r_{GC} = 1$ pc will then be the relatively long viscous time
of the disk, and the fate of much of the disk will be to migrate inward and
eventually lead to direct
accretion onto the SMBH plus a concomitant outflow of material energized by the
accretion.  

\section{Conclusions}

We performed simulations of SN shells expanding inside the NSC near the SMBH with the code RING using the infinitesimally thin shell approximation in 3D.
In our model, the homogeneously distributed ISM rotates following the rotation curve (\ref{eq-rotation}), which partially balances the gravitational attraction of the NSC and SMBH. We tested three different ISM densities, $n_{out} = 10^3, 10^4$, and $10^5$ cm$^{-3}$, and identified the delivery region from where the SN expanding shells deliver mass into the central 1 pc. It is a region of small angular momentum with a low  value of $R_L$ centered around the rotational axis (see  Eq. (\ref{Ram})).

The simulations  show that SNe occurring within a conical region around the rotational axis of the galaxy can feed the central accretion disk surrounding the SMBH. A burst forming $10^5$ M$_\odot$ of stars within the central 100 pc will produce $\sim 1000$ SNe during the subsequent 30 Myr, which will deliver with their expanding shells a total mass of $(1 - 2) \times 10^5$ M$_\odot$ to the central parsec,  for ISM density $n_{out} = 10^3$ cm$^{-3}$, or $(9 \times 10^3 - 10^6)$ M$_\odot$ for ISM density $n_{out} = 10^5$ cm$^{-3}$. The corresponding continuous SFR = $\rm 0.003$ M$_\odot$ /yr leads to an average mass flux into the central parsec  of $\rm (3 - 7) \times 10^{-3}$ or  $\rm (0.3 - 40) \times 10^{-3}$ M$_\odot$ /yr. A high SFR $\rm \sim 30$ M$_\odot$ /yr  inside the central 100 pc of the galaxy will increase this mass inflow rate to $\rm \le 3 - 400$ M$_\odot$ /yr depending on the ISM density and on the radial distribution of SNe.

\begin{acknowledgements}
We are grateful to the anonymous referee whose constructive comments improved the manuscript.
  This study has been supported by Czech Science Foundation Grant 19-15480S
  and by the project RVO:67985815.
\end{acknowledgements}

%
%

\bibliographystyle{aa} 
\bibliography{palous-etal-38768.bib} 

\end{document}